\documentclass[11pt,twoside]{article}
\usepackage[pdftex]{graphicx}
\usepackage{amsmath}
\usepackage{amssymb}
\usepackage{cite}
\begin{document}
\newcommand{\pst}{\hspace*{1.5em}}


\newcommand{\be}{\begin{equation}}
\newcommand{\ee}{\end{equation}}
\newcommand{\bm}{\boldmath}
\newcommand{\ds}{\displaystyle}
\newcommand{\bea}{\begin{eqnarray}}
\newcommand{\eea}{\end{eqnarray}}
\newcommand{\ba}{\begin{array}}
\newcommand{\ea}{\end{array}}
\newcommand{\arcsinh}{\mathop{\rm arcsinh}\nolimits}
\newcommand{\arctanh}{\mathop{\rm arctanh}\nolimits}
\newcommand{\bc}{\begin{center}}
\newcommand{\ec}{\end{center}}

\thispagestyle{plain}

\label{sh}


\begin{center} {\Large \bf
\begin{tabular}{c}
Subadditivity condition for spin-tomograms and density matrices\\ of arbitrary composite
and noncomposite qudit systems
\end{tabular}
}
\end{center}

\bigskip

\bigskip

\begin{center} {\bf
V. N. Chernega, O. V. Man'ko$^*$, V. I. Man'ko }
\end{center}

\medskip

\begin{center}
{\it P.~N.~Lebedev Physical Institute, Russian Academy of Sciences\\
Leninskii Prospect 53, Moscow 119991, Russia}
\end{center}

\smallskip

\smallskip

$^*$Corresponding author e-mail:~~~omanko@sci.lebedev.ru

\begin{abstract}\noindent
New quantum entropic inequality for states of system of $n\geq 1$ qudits is obtained. The inequality has the form of quantum subadditivity condition of bipartite qudit system and coincides with this subadditivity condition for the system of two qudits. The general statement on existence of the subadditivity condition for arbitrary probability distribution and arbitrary qudit-system tomogram is formulated. The nonlinear quantum channels creating the entangled states from separable ones are discussed.
\end{abstract}

\medskip

\noindent{\bf Keywords:} entropy, information, tomographic probability, qubits, qudit, subadditivity condition, nonlinear quantum channels.

\section{Introduction}
\pst The probability distributions are characterized by Shannon entropy \cite{Shanon}. The state of quantum systems, identified with density matrices \cite{Landau,Landau1,vonNeuman,vonNeuman1} are characterized by von Neumann entropy. For the pure states identified with the wave functions the von Neumann entropy is equal to zero. The entropies correspond to order in the system \cite{Holevo}. For complete order in the classical system the Shannon entropy equals to zero. For composite classical and quantum systems there exist some inequalities related to the entropies of the system and its subsystems. The inequalities for von Neumann entropies of bipartite quantum system mean that the sum of the entropies of the subsystems is larger or equal to the entropy of the composite system. Analogous inequality holds for Shannon entropy \cite{Shanon} of the bipartite system. Recently, \cite{Mancini96,OlgaJRLR1997,DodPLA,OgaJETP,OlgaBregence} it was shown that the quantum states can be identified with tomographic probability distributions called quantum tomograms both for discrete spin (qudit) states and for the systems with the continious variables like system of interacting oscillators. In view of this the inequalities known for classical probability distributions can be obtained also for quantum tomograms \cite{FoundPhysRita,Fedele,Vaxsha,Turin,ChernegaJRLRv32,OlgaVova}. Recent review of probability vector properties both in classical and quantum domains is presented in \cite{JRLRN1Maromo2014}. Recently, it was clarified \cite{mamapapa} that the inequalities like subadditivity condition known for bipartite system can be found also for noncomposite systems.
The idea of this approach is based on qubit portrait method of qudit states suggested in \cite{Vovf} and applied to study entanglement properties of bipartite qudit systems in \cite{Lupo}. There exist \cite{LiebSeiringer} some inequalities for von Neumann entropy of bipartite system connecting "classical" and quantum entropies. The aim of our  work is to use the approach of extending the inequalities known for composite systems considered in \cite{Vovacompositesystems} and to obtain new inequalities for tomographic entropies  both for composite and noncomposite quantum systems. The model of quantum mechanics based on classical gaussian probability distribution is elaborated in \cite{Khrennikov,Khrennikov1,Khrennikov2}.

The paper is organized as follows. In second section we discuss probability vectors and entropic inequality for bipartite system. In third section we generalized subadditivity condition for arbitrary probability vector ${\bf P}$. In fourth section we review the method of the portrait of density matrices and in fifth section we discuss, as an example, system states with density $6\times6$-matrices. In sixth section we discuss nonlinear chains of maps of probability vectors. The conclusions and perspectives are given in seventh section.

\section{Probability vectors and entropic inequality for bipartite system}
Let us consider a set of $N$ nonnegative numbers $p_1,p_2,\ldots,p_N$ such that $\sum_{k=1}^{N}p_k=1$. The set of the numbers can be identified with a probability vector ${\bf P}=(p_1,p_2,\ldots,p_N$) where the numbers $p_k$ $(k=1,2,\ldots,N)$ are related to the results of measuring a system random variable. The variable is assumed to give $N$ different values. These numbers $p_k$ provide the probability to get $k^{th}$ value of the random variable.
For systems of qudits the components of probability vector  ${\bf P}$ can be identified with the $n$ values of qudit state tomograms $w({\bf m},u)=\langle{\bf m}|u\rho u^+|{\bf m}\rangle$, where $\rho$ is density matrix, $u$ is unitary matrix and vector ${\bf m}=({\bf m}_1,{\bf m}_2,\ldots,{\bf m}_n)$ with ${\bf m}_k=(-j_k,-j_k+1,\ldots,j_k)$ being spin $j_k$ projection. If one considers the system which contains two subsystems (bipartite system) the measuring the values of two random variables gives the table of $n=N\cdot M$ nonnegative numbers $p_{kj}$ $(k=1,2,\ldots,N,\, j=1,2,\ldots,M)$. The numbers provide joint probability distribution associated with the results of measuring two random variables. The joint probability distribution is normalised, i.e.
\begin{equation}\label{eq.1}
  \sum_{k=1}^N\sum_{j=1}^M p_{kj}=1.
\end{equation}
If one measures only one of these two random variables the joint probability distribution determines the marginal probability distribution
\begin{equation}\label{eq.2}
{\cal P}_k=\sum_{j=1}^M p_{kj},\quad\sum_{k=1}^N{\cal P}_k=1.
\end{equation}
Another marginal probability distribution describing the results of measuring the second random variable reads
\begin{equation}\label{eq.3}
\Pi_j=\sum_{k=1}^N p_{kj}, \quad\sum_{j=1}^M\Pi_j=1.
\end{equation}
If the random variables are independent (there is no correlations between the subsystems of the bipartite system) the numbers $p_{kj}$ have the factorized form
\begin{equation}\label{eq.4}
  p_{kj}={\cal P}_k\Pi_j.
\end{equation}
The marginal distributions can be associated with two probability vectors $\vec{\cal P}=({\cal P}_1,{\cal P}_2,\ldots,{\cal P}_N)$ and ${\bf\Pi}=(\Pi_1,\Pi_2,\ldots,\Pi_M)$. The table of numbers $p_{kj}$ also can be described by the probability vector ${\bf P}$. In fact any column vector can be considered as rectangular matrix. Then the vector (rectangular matrix) ${\bf P}$ is expressed in terms of two rectangular matrices (vector $ \vec{\cal P}$ and ${\bf\Pi}$) as their direct product
\begin{equation}\label{eq.5}
  {\bf P}=\vec{\cal P}\otimes{\bf \Pi},\quad {\bf P}=(P_1,P_2,\quad\ldots,\quad P_{NM}).
\end{equation}
It means that we use the invertable map of natural numbers onto pairs of integers $n\Longleftrightarrow(kj)$ which explicitly reads
\begin{equation}\label{eq.6}
1\Longleftrightarrow 11,\quad 2\Longleftrightarrow 21,\quad\ldots,\quad N\Longleftrightarrow N1,\quad N+1\Longleftrightarrow 21,\quad \ldots,\quad n\Longleftrightarrow N\cdot M.
\end{equation}
In fact we code the natural numbers $1,2,\,\ldots, \,n=N\cdot M$ by pairs of the natural numbers $(k j)$ where $k=1,2,\ldots,\,N$,  $j=1,2,\ldots,\,M$. Let us for simplicity assume that $N\leq M$. Any probability distribution is characterized by Shannon entropy \cite{Shanon}. For example the joint probability distribution $p_{kj}$ for bipartite system has the Shannon entropy $H(1,2)$ determined as
\begin{equation}\label{eq.7}
H(1,2)=-\sum_{k=1}^{N}\sum_{j=1}^{M} p_{kj}\ln\,p_{kj}.
\end{equation}
The marginal probability distributions have the Shannon entropies $H(1)$ and $H(2)$ of the form
\begin{equation}\label{eq.7}
H(1)=-\sum_{k=1}^N{\cal P}_k\,\ln{\cal P}_k,\quad H(2)=-\sum_{j=1}^M\Pi_j\,\ln\Pi_j.
\end{equation}
It is worthy to note that the entropy $H(1,2)$ can be written in the form
\begin{equation}\label{eq.8}
H\equiv H(1,2)=-\sum_{n=1}^{N M} P_n\ln\,P_n.
\end{equation}
For all these entropies we introduce the vector notations. The entropy
\begin{equation}\label{eq.10}
H=-{\bf P}\ln{\bf P}, \quad H(1)=-\vec{\cal P}\,\ln\,\vec{\cal P}, \quad H(2)=-{\bf \Pi}\,\ln{\bf \Pi}.
\end{equation}
In the formula (\ref{eq.10}) we used the following definition:
\[{\bf x}\ln{\bf x}=\equiv\sum_{\alpha=1}^L x_{\alpha}\ln x_\alpha.\]
It means that ${\bf x}=(x_1,x_2,\ldots,x_L)$, and $\alpha=1,2,\ldots,L.$ Using the vector notations gives the possibility to describe the Shannon entropy of bipartite system with two random variables associated with the joint probability distribution $p_{k j}$ and the system with one random variable associated with the probability distribution $p_k$ by identical formulas presented in (\ref{eq.10}). The only difference between the expressions $H,\,H(1)$ and $H(2)$ is that the "scalar product" in (\ref{eq.10}) is evaluated for the vectors which have different number of components. In (\ref{eq.10}) the vector $\vec{\cal P}$ has $N$ components, the vector ${\bf \Pi}$ has $M$ components and the vector ${\bf P}$ has $n=N\cdot M $ components. This difference can be removed. In fact, since
$\lim_{x\rightarrow 0}x\ln x=0$ we can consider vectors $\vec{\cal P}$ and ${\bf \Pi}$ as vectors with $n=N M$ components by adding the zero components to the initial vectors, i.e.
\begin{equation}\label{eq.11}
\vec{\cal P}=({\cal P}_1,{\cal P}_2,\ldots,{\cal P}_N,0,0,\ldots,{\cal P}_{M N}=0),
\end{equation}
\begin{equation}\label{eq.12}
{\bf \Pi}=( \Pi_1, \Pi_2,\ldots, \Pi_M,0,0,\ldots, \Pi_{N M}=0).
\end{equation}
Using these new vectors we do not change the values of the entropies, i.e. in formulas (\ref{eq.10}) we have the same expressions but all the probability--vectors  ${\bf P}$, $\vec{\cal P}$ and ${\bf \Pi}$ are considered as vectors with $n=N M$ components. It is known that the marginals ${\cal P}_k$ and $\Pi_j$ of joint probability distribution $p_{k j}$ satisfy the entropic inequality called subadditivity condition which reads
\begin{equation}\label{eq.13}
H(1,2)\leq H(1)+H(2),
\end{equation}
where the Shannon entropies are given by (\ref{eq.6})-(\ref{eq.8}). In explicit form  this inequality reads
\begin{equation}\label{eq.14}
-\sum_{k=1}^N\sum_{j=1}^M p_{k j}\ln p_{k j}\leq-\sum_{k=1}^N{\cal P}_k\ln{\cal P}_k-\sum_{j=1}^M\Pi_j\ln\Pi_j.
\end{equation}
For the case of independent random variables $p_{k j}={\cal P}_k\Pi_j$ one has equality
\begin{equation}\label{eq.15}
H(1,2)= H(1)+H(2).
\end{equation}
The Shannon mutual information is defined as the difference of entropies
\begin{equation}\label{eq.16}
I= H(1)+H(2)-H(1,2).
\end{equation}
This information satisfies the nonnegativity condition $I\geq0$.

Using vector notations we can write the subadditivity condition (\ref{eq.14}) in the form
\begin{equation}\label{eq.17}
-{\bf P}\ln{\bf P}\leq -\vec{\cal P}\ln\vec{\cal P}-{\bf \Pi}\ln{\bf \Pi},
\end{equation}
where all the probability--vectors have $N\cdot M$ components.

The Shannon information is expressed in terms of the probability $n$-vectors as
\begin{equation}\label{eq.18}
I= -\vec{\cal P}\ln\vec{\cal P}-{\bf \Pi}\ln{\bf \Pi}+{\bf P}\ln{\bf P}.
\end{equation}
Here $n=N\cdot M$.

\section{Generalization of subadditivity condition\\ for arbitrary probability vector ${\bf P}$}
The subadditivity condition (\ref{eq.17}) written as inequality for three probability $n$-vectors ${\bf P}$, $\vec{\cal P}$ and ${\bf \Pi}$ provides the possibility to generalyze the inequality and to prove that such inequality takes place for arbitrary probability $n$-vectors. To clarify this issue let us express the $n$-vectors $\vec{\cal P}$ and ${\bf \Pi}$ in terms of two stochastic $n\times n$-matrices $M_{12}$ and $M_{21}$ and vector ${\bf P}$.

In fact one can observe that the following equalities hold
\begin{equation}\label{eq.19}
\vec{\cal P}=M_{12}{\bf P},\quad {\bf \Pi}=M_{21}{\bf P},
\end{equation}
where the stochastic matrices $M_{12}$ and $M_{21}$ read
\begin{equation}\label{eq.20}
M_{12}=\left(
  \begin{array}{cccc}
    1_M & 0_M & \ldots & 0_M \\
    0_M & 1_M & \ldots & 0_M \\
    \ldots & \ldots & \ldots & \ldots \\
    0_M & 0_M & \ldots & 1_M \\
    \cdot & 0_S & \cdot & \cdot \\
  \end{array}
\right), \quad
M_{21}=\left(
  \begin{array}{cccc}
    1_N & 0_N & \ldots & 0_N \\
    0_N & 1_N & \ldots & 0_N \\
    \ldots & \ldots & \ldots & \ldots \\
    0_N & 0_N & \ldots & 1_N \\
    \cdot & 0_Q & \ldots & \cdot \\
  \end{array}
\right).
\end{equation}
Here the rectangular matrices $1_M$ and $0_M$ with one row and $M$ columns read
\begin{equation}\label{eq.21}
1_M=(1,1,\ldots,1), \quad 0_M=(0,0,\ldots,0).
\end{equation}
The zero rectangular matrix $0_S$ has $N\cdot M-N$ rows and $N\cdot M$ columns. The $N\times N$ - blocks in the matrix $M_{21}$ are the unity $N\times N$-matrix $1_N$ and zero $N\times N$-matrix $0_N$. The zero matrix $0_Q$ contains $N\cdot M-M$ rows and $N\cdot M$ columns. Using formula (\ref{eq.19}) we can rewrite subadditivity condition (\ref{eq.17}) known for joint probability distribution of bipartite system in the form
\begin{equation}\label{eq.22}
-{\bf P}\ln{\bf P}\leq -(M_{12}{\bf P})\ln(M_{12}{\bf P})-(M_{21}{\bf P})\ln(M_{21}{\bf P}).
\end{equation}
We get the inequality (\ref{eq.22}) as the property of joint probability distribution of bipartite system. But it is obvious that this inequality is the inequality which is valid for arbitrary set of $n=N\cdot M$ nonnegative numbers $(p_1,p_2,\ldots,p_{N M})$. In view of this one can formulate the general statement: Given arbitrary probability vector ${\bf P}$ with $n$ components where the integer $n$ can be presented in the product form of two integers $n=N\cdot M$, $N\leq M$. Then the inequality (\ref{eq.22}) holds where the matrices (\ref{eq.20}) are two stochastic matrices containing only zeros and unities. The inequality (\ref{eq.22}) is valid also for all $n!$ vectors ${\bf P}_{per}$ obtained from the initial vector ${\bf P}$ by means of permutations of the indices $(1,2,\ldots,n)$ labeling the vector components. It means
\begin{equation}\label{eq.23}
-{\bf P}\ln{\bf P}=-{\bf P}_{per}\ln{\bf P}_{per}\leq -(M_{12}{\bf P}_{per})\ln (M_{12}{\bf P}_{per})-(M_{21}{\bf P}_{per})\ln(M_{21}{\bf P}_{per})
\end{equation}
It is worthy to note that the integer $n$ can have different product decomposition $n=\bar N\bar M$. The equality (\ref{eq.22}) and (\ref{eq.23}) take place also with new matrices $\bar M_{12}, \bar M_{21}$ obtained from (\ref{eq.20}) by the substitution $N\rightarrow\bar N$ and  $M\rightarrow\bar M$. It is worthy to note that the inequality (\ref{eq.22}) holds for arbitrary probability vector ${\bf P}$ which corresponds to a point on simplex including the vectors which have some zero components. We use the remark to extend our inequality (\ref{eq.22}) for arbitrary probability $n$-vectors including the case of prime number $n$. To write the inequality for such probability $n$-vector ${\bf P}$ we construct new vector ${\bf P}'=(p_1,p_2,\ldots,p_n,0,0,,\ldots,p_{n'}=0)$. The $n'$-vector ${\bf P}'$ has $n'$ components. We added the appropriate quantity of zero components to the initial $n$-vector ${\bf P}$ such that the new integer $n'$ has the product form $n'=N'M'$. It is clear that there are many ways to construct such vectors with different integers $n'\geq n$. All these vectors will satisfy the subadditivity condition. Another generalisation of obtained inequality can be formulated for arbitrary set of nonnegative numbers $x_1,x_2,\ldots,x_n$. These numbers correspond to a point on the cone. Using the map
\[x_k\rightarrow p_k=\frac{x_k}{\sum_{j=1}^n x_j}\]
and applying the inequality (\ref{eq.22}) to the vector
\[\frac{{\bf x}}{\sum_{j=1}^n x_j}={\bf P}\]
we get inequality for arbitrary finite set of $n$ nonnegative numbers $x_k$, i.e.
\begin{equation}\label{eq.24}
-{\bf x}\ln{\bf x}\leq -(M_{12}{\bf x})\ln(M_{12}{\bf x})-(M_{21}{\bf x})\ln(M_{21}{\bf x})+(\sum_{j=1}^n x_j)\ln(\sum_{j=1}^n x_j).
\end{equation}
Thus we proved that the coordinates of a point on a cone satisfy the analog of subadditivity condition with extra terms in the right-hand side of (\ref{eq.24}). For arbitrary integers $n$ the stohastic matrices $M_{12}$ and $M_{21}$ can be written in fixed canonical form . We can introduce the information on the cone which is the difference of the right hand side and left-hand side of Eq.(\ref{eq.24}), i.e.
\begin{equation}\label{eq.25}
I_{\bf x}=-(M_{12}{\bf x})\ln(M_{12}{\bf x})-(M_{21}{\bf x})\ln(M_{21}{\bf x})+{\bf x}\ln{\bf x}+(\sum_{j=1}^n x_j)\ln(\sum_{j=1}^n x_j).
\end{equation}
If $\sum_{j=1}^n x_j=1$ we have the point on the simplex and the information $I_{{\bf x} }$ becomes the analog of Shannon information which we introduced for arbitrary probability distribution described by a probability vector ${\bf P}$. It reads
\begin{equation}\label{eq.26}
I_{\bf p}=-(M_{12}{\bf P})\ln(M_{12}{\bf P})-(M_{21}{\bf P})\ln(M_{21}{\bf P})+{\bf P}\ln{\bf P}\geq0.
\end{equation}
There exist $n!$ informations $I_{{\bf p}}$ obtained from (\ref{eq.26}) by replacing probability vector ${\bf P}\rightarrow {\bf P}_{per}.$ In case of bipartite systems and $n=N\cdot M$, where $N$ and $M$ correspond to outcomes of two random variables the information $I_{\bf p}$ coincides with Shannon mutual information. The meaning of introduced informations $I_{{\bf p}_{per}}$ and informations (\ref{eq.26}) introduced for arbitrary probability vector ${\bf P}$ needs the additional clarification.

\section{Portrait of density matrices}
We apply the analogous method to get the positive map of $n\times n$ density matrix $\rho(1,2)$ of bipartite system with $n=N\cdot M$, $N\leq M$. In fact, if the state $\rho(1,2)$ is simply separable state, i.e. $\rho(1,2)=\rho(1)\otimes\rho(2)$ and $\rho(1)$ is $N\times N$-matrix, $\rho(2)$ is $M\times M$-matrix we can observe that the $N\times N$ matrix $\rho(1)$ is given by the following procedure. Namely, the matrix elements $\rho_{k j}(1)$, $k,j,=1,2,\ldots,N$ are given as the first $N$ vector components of $N\cdot M$ vectors $\vec\rho_1(1),\vec\rho_2(1),\ldots,\vec\rho_N(1)$ where
\begin{equation}\label{100}
\vec\rho_1(1)=M_{12} {\bf R}_1, \quad \vec\rho_2(1)=M_{12} {\bf R}_2, \quad\dots,\quad \vec\rho_N(1)=M_{12} {\bf R}_N.
\end{equation}
Here the $N\cdot M$ matrix $M_{12}$ is given by Eq.(\ref{eq.20}). The $N\cdot M$ vectors ${\bf R}_j$, $j=1,2,\ldots N$ have the components
\begin{equation}\label{101}
 ({\bf R}_j)_{k\alpha}=\rho_{k j}(1)\rho_{\alpha\alpha}(2), \quad k=1,2,\ldots,N, \,\alpha=1,2,\ldots,M.
\end{equation}
Thus we used the invertable map of integers $1,2,\ldots,N$, $\alpha=1,\ldots,M$ onto the pairs of integers $k,\alpha$ $1\leftrightarrow 11, \,2\leftrightarrow 21,\ldots,\,n \leftrightarrow N M$ to label the components of the vector ${\bf R}_j$.

If the matrix $\rho(1,2)$ had the generic form with matrix elements $\rho_{k\alpha j\beta}(1,2)$ we have the positive map $\rho(1,2)\rightarrow\rho(1)$ given by the same formula (\ref{100}) with changed vectors ${\bf R}_j$. Namely the vectors ${\bf R}_j$ have the components
\begin{equation}\label{103}
({\bf R}_j)_{k \alpha}=\rho_{k\alpha\,j\alpha}(1,2).
\end{equation}
There is no sum over indices $\alpha$. Thus the described construction provides the map of the $N\cdot M$ density matrix $\rho(1,2)$ onto the density matrix which also can be considered as $N\cdot M$ matrix $\bar \rho(1)$ of the form
\begin{equation}\label{104}
\bar\rho(1)=\left(
              \begin{array}{cc}
                \rho(1) & 0_1 \\
                0_1^{tr} & 0_{n-N} \\
              \end{array}
            \right).
\end{equation}
Here $0_1$ is zero rectangular matrix with $N$ rows and $n-N$ columns, the matrix $0_{n-N}$, where $n=N\cdot M$ has the zero matrix elements. Analogous construction can be applied to get the map $\rho(1,2)\rightarrow\rho(2)=\mbox{Tr}_1\rho(1,2)$. The explicit form of this map can be obtained from (\ref{104}) by using the known matrix of map of vectors ${\bf a}\otimes{\bf b}\longleftrightarrow{\bf b}\otimes {\bf a}$ given by the matrix $S$ such that
\begin{equation}\label{105}
({\bf a}\otimes{\bf b})_k=\sum_{m=1}^n S_{km}({\bf b}\otimes{\bf a})_m.
\end{equation}
Using the matrix $S$ we can reduce the problem of finding the expression  for the matrix $\rho(2)$ to the problem discussed above with the replacement discussed above $1\leftrightarrow 2,\, N\leftrightarrow M$. The $N\cdot M$ matrices $\bar\rho(1)$ and $\bar\rho(2)$ satisfy the subadditivity condition
\begin{equation}\label{105a}
-\mbox{Tr}\bar\rho(1)\ln\bar\rho(1)-\mbox{Tr}\bar\rho(2)\ln\bar\rho(2)\geq-\mbox{Tr}\bar\rho(1,2)\ln\bar\rho(1,2).
\end{equation}
Thus for arbitrary $n\times n$ matrix $\rho$, where $n=N\cdot M$ we can obtain two matrices $\rho(1)$ and $\rho(2)$ applying to the initial matrix $\rho$ the map which naturally can be applied to the bipartite matrix $\rho(1,2)$. The matrix $\rho$ can be considered as density matrix of one qudit only. Nevertheless the associated with it matrices $\bar\rho(1)$ and $\bar\rho(2)$, satisfy the subadditivity condition (\ref{105a}).

\section{Example of system states with $6\times6$ - matrices}
To demonbstrate our approach let us consider the example of $n=6$. We can consider the density matrix $\rho_{k j}$, $k,\,j\,=1,2,\ldots,6$ as the density matrix of one qudit with $j=5/2$. The matrix $\rho$ reads
\begin{equation}\label{106}
\left(
  \begin{array}{cccccc}
    \rho_{11} &\rho_{12}  &\rho_{13}  &\rho_{14}  & \rho_{15} & \rho_{16} \\
    \rho_{21} &\rho_{22} &\rho_{23}  & \rho_{24} &\rho_{25} &\rho_{26} \\
    \rho_{31} &\rho_{32}  & \rho_{33} & \rho_{34} & \rho_{35} &\rho_{36} \\
    \rho_{41} & \rho_{42} &\rho_{43}  &\rho_{44} &\rho_{45} &\rho_{46} \\
    \rho_{51} &\rho_{52}  &\rho_{53} &\rho_{54}  &\rho_{55} &\rho_{56} \\
    \rho_{61} &\rho_{62} &\rho_{63}  & \rho_{64} &\rho_{65}  &\rho_{66} \\
  \end{array}
\right)\equiv\left(
               \begin{array}{cc}
                 \rho^{(1)}&\rho^{(2)} \\
                 \rho^{(3)} &\rho^{(4)} \\
               \end{array}
             \right).
\end{equation}
Here matrices $\rho^{(k)},\, k=1,2,3,4$ are $3\times3$-matrices which constitute $\rho$. Let us take integers $N=2,\, M=3$. The $6\times6$ stochastic matrix $M_{12}$ reads
\begin{equation}\label{107}
M_{12}=\left(
  \begin{array}{cccccc}
  1 & 1 & 1 & 0 & 0 & 0 \\
0 & 0 &0 & 1 & 1 & 1 \\
   0 & 0 & 0 & 0 & 0 & 0 \\
    0 & 0 & 0 & 0 & 0 & 0 \\
    0 & 0 & 0 & 0 & 0 & 0 \\
  \end{array}
\right).
\end{equation}
The $6$-vectors ${\bf R}_1, \, {\bf R}_2$ read
\begin{equation}\label{108}
{\bf R}_1=\left(
  \begin{array}{c}
    \rho_{11} \\
    \rho_{22} \\
   \rho_{33}  \\
    \rho_{41} \\
    \rho_{52} \\
   \rho_{63} \\
  \end{array}
\right),\quad {\bf R}_2=\left(
  \begin{array}{c}
    \rho_{14} \\
   \rho_{25} \\
   \rho_{36}  \\
   \rho_{44}  \\
   \rho_{55} \\
    \rho_{66} \\
  \end{array}
\right).
\end{equation}
Applying the matrix $M_{12}$ to vectors (\ref{108}) we get $2\times2$ - matrix $\rho(1)$ of the form
\begin{equation}\label{109}
\rho(1)=\left(
          \begin{array}{cc}
            \rho_{11}+\rho_{22}+\rho_{33} &\rho_{14}+\rho_{25}+\rho_{36} \\
            \rho_{41}+\rho_{52}+\rho_{63} &\rho_{44}+\rho_{55}+\rho_{66} \\
          \end{array}
        \right)
\end{equation}
The $6\times6$ - matrix $\bar\rho(1)$ has the form
\begin{equation}\label{110}
\bar\rho(1)=\left(
  \begin{array}{cc}
    \rho(1) &0_{24} \\
    0_{24}^{tr} &0_4 \\
  \end{array}
\right),\quad 0_{24}=\left(
                       \begin{array}{cccc}
                         0 & 0 & 0 & 0 \\
                         0 & 0 & 0 & 0 \\
                       \end{array}
                     \right)
\end{equation}
and $0_4$ is $4\times4$ - matrix with zero matrix elements. The matrix $\bar\rho(2)$ has the form
\begin{equation}\label{111}
\bar\rho(2)=\left(
              \begin{array}{cc}
                \rho(2) & 0 \\
                0 & 0 \\
              \end{array}
            \right), \quad \rho(2)=\rho^{(1)}+\rho^{(4)}.
\end{equation}
In general case of $n\times n$ - matrix $\rho$ one has analogous map $\rho\rightarrow\bar\rho(1)$, $\rho\rightarrow\bar\rho(2)$. The $M\times M$ - matrix $\rho(2)$ equals to the sum of $N$ blocks of the matrix $\rho$, i.e.
\begin{equation}\label{112}
\rho(2)=\sum_{k=1}^N\rho^{(k)}.
\end{equation}
Each block $\rho^{(k)}$ is the $M\times M$ - matrix. These $N$ blocks constitute the density matrix of the state which is obtained by "decoherence" map from the initial matrix $\rho$. Namely we construct from $\rho$ the block--diagonal matrix
\[\rho_d=\left(
          \begin{array}{cccc}
            \rho^{(1)} & 0 & 0 \\
            0 & \rho^{(2)} & 0 \\
            \ldots & \ldots & \ldots \\
            0 & 0& \rho^{(N)} \\
          \end{array}
        \right)\]
keeping $N$ of the $M\times M$ - matrices and other matrix elements assume to be equal zero. Then we sum all these blocks. As result we get matrix $\rho(2)$.  So, the matrix $\bar\rho(2)$ is a "portrait" of the initial matrix $\rho$. Another "portrait" is the matrix $\bar\rho(1)$ which for initial $6\times6$ - matrix $\rho$ is given by Eq.(\ref{110}). There is possibility to make another map of the matrix $\rho$ onto two "portrait" matrices $\bar\rho(1)$ and $\bar\rho(2)$, namely we take $N=3$ and $M=2$. Then the $3\times3$ - matrix $\rho(1)$ reads
\begin{equation}\label{113}
\rho(1)=\left(
          \begin{array}{ccc}
            \rho_{11}+\rho_{22} &\rho_{13}+\rho_{24} &\rho_{15}+\rho_{26} \\
            \rho_{31}+\rho_{42} &\rho_{33}+\rho_{44} &\rho_{35}+\rho_{46} \\
            \rho_{51}+\rho_{62} &\rho_{53}+\rho_{64} &\rho_{55}+\rho_{66} \\
          \end{array}
        \right)
\end{equation}
and the $2\times 2$ - matrix $\rho(2)$ is
\begin{equation}\label{114}
\rho(2)=\left(
  \begin{array}{cc}
    \rho_{11}+\rho_{33}+\rho_{55} &\rho_{12}+\rho_{34}+\rho_{56} \\
    \rho_{21}+\rho_{43}+\rho_{65}&\rho_{22}+\rho_{44}+\rho_{66} \\
  \end{array}
\right).
\end{equation}
The subadditivity inequality for all the obtained pairs $\rho(1)$ and $\rho(2)$ (i.e. $\bar\rho(1)$, $\bar\rho(2)$) is given as
\begin{equation}\label{115}
-\mbox{Tr}(\rho(1)\ln\rho(1))-\mbox{Tr}(\rho(2)\ln\rho(2))=-\mbox{Tr}(\bar\rho(1)\ln\bar\rho(1))
-\mbox{Tr}(\bar\rho(2)\ln\bar\rho(2))
\geq-\mbox{Tr}(\rho\ln\rho).
\end{equation}
The von Neumann quantum mutual information is given by difference
\begin{equation}\label{116}
I_{q}(\bar\rho(1),\bar\rho(2))=-\mbox{Tr}(\bar\rho(1)\ln\bar\rho(1))-\mbox{Tr}(\bar\rho(2)\ln\bar\rho(2))
+\mbox{Tr}(\rho\ln\rho).
\end{equation}
The inequalities for entropies (\ref{115}) are valid for $(N M)!$ matrices $\bar\rho(1)$, $\bar\rho(2)$ obtained by means of all permutations of integers $1,2,\ldots,n\rightarrow1_p,2_p,\ldots,n_p$, determining matrix elements of the $n\times n$ - matrix $\rho$. It is worthy to note that if one has any $n\times n$ - density matrix $\rho$ one cen construct the matrix $\rho_{n'}$, where $n'=n+p= N\cdot M$ which reads
\begin{equation}\label{117}
\left(
  \begin{array}{cc}
    \rho & 0 \\
    0 & 0 \\
  \end{array}
\right).
\end{equation}
After this one obtains by described procedure all the maps $\rho_{n'}\rightarrow\bar\rho^{(n')}(1)$ and $\bar\rho^{(n')}(2)$ and the new subadditivity conditions are written for these matrices
\begin{equation}\label{118}
-\mbox{Tr}(\bar\rho^{(n')}(1)\ln\bar\rho^{(n')}(1))-\mbox{Tr}(\bar\rho^{(n')}(2)\ln\bar\rho^{(n')}(2))\geq
-\mbox{Tr}(\rho^{(n')}\ln\rho^{(n')})=-\mbox{Tr}(\rho\ln\rho).
\end{equation}
Analogously the nonnegative mutual informations are given by the difference
\begin{equation}\label{119}
I_{\rho}'(\rho^{(n')}(1)),\rho^{(n')}(2))=-\mbox{Tr}(\bar\rho^{(n')}(1)\ln\bar\rho^{(n')}(1))
-\mbox{Tr}(\bar\rho^{(n')}(2)\ln\bar\rho^{(n')}(2))+\mbox{Tr}(\rho\ln\rho).
\end{equation}
These informations depend on the maps of the matrix $\rho\rightarrow\rho^{(n')}(1)$, $\rho\rightarrow\rho^{(n')}(2)$.
For example if one has the $5\times5$ - density matrix $\rho$ corresponding to qudit with $j=2$ (i.e. $n=5$ and one can take $n'=n+1=6$) the pairs of matrices $\rho(1)$ and $\rho(2)$  obtained by the described positive maps are, e.g.
\begin{equation}\label{116a}
\rho(1)=\left(
          \begin{array}{ccc}
            \rho_{11}+\rho_{22} &\rho_{13}+\rho_{24} &\rho_{15} \\
            \rho_{31}+\rho_{42} &\rho_{33}+\rho_{44} &\rho_{35} \\
            \rho_{51} &\rho_{53} &\rho_{55} \\
          \end{array}
        \right), \quad
\rho(2)=\left(
          \begin{array}{cc}
            \rho_{11}+\rho_{33}+\rho_{55} &\rho_{12}+\rho_{34} \\
            \rho_{24}+\rho_{43} &\rho_{22}+\rho_{44} \\
          \end{array}
        \right).
\end{equation}
One has
\begin{equation}\label{117a}
-\mbox{Tr}(\rho(1)\ln\rho(1))-\mbox{Tr}(\rho(2)\ln\rho(2))\geq -\mbox{Tr}(\rho\ln\rho).
\end{equation}
Other pairs are also obtained by means of coding the pairs for generic $6\times6$ - density matrix $\rho_{k j}$ and assuming that all the matrix elements $\rho_{k6}$ and $\rho_{6 j}$ equal to zero. Then all the subadditivity conditions for qudit $j=2$ states are obtained from constructed one by permutations of the integers $1,2,3,4,5\mapsto1_p,2_p,3_p,4_p,5_p$ labeling matrix elements of the matrix $\rho_{k j},$ $(k,j=1,2,3,4,5).$

\section{Nonlinear maps of probability vectors}
Let us discuss the possibility to make a general map of probability vector ${\bf p}=(p_1,p_2,\ldots,p_n)$ onto probability vector $\vec\Pi=(\Pi_1({\bf p}),\Pi_2({\bf p}),\ldots,\Pi_m({\bf p}))$ and the vector components of vector $\vec\Pi$, i.e. $\Pi_k({\bf p})$ are some functions of the vector ${\bf p}$. If $n=m$ and for particular case of linear functions we have the form of the map
\begin{equation}\label{eqq1}
\vec\Pi({\bf p})=M{\bf p},
\end{equation}
where the $n\times n$ - matrix $M$ has the matrix elements with the property $\sum_{k=1}^n M_{k j}=1$. If Eq.(\ref{eqq1}) provides the linear map for all the vectors ${\bf p}$ belonging to symplex the matrices $M$ are stochastic matrices with nonnegative matrix elements. If Eq.(\ref{eqq1}) provides the linear map for vector belonging to some domain in the symplex the matrices $M$ can have negative matrix elements. In all these cases the matrices $M$ form a semigroup. In particular, the stochastic matrices form the semigroup. One  can introduce nonlinear maps of the probability vectors, choosing the specific functions $\Pi_k({\bf p})$ which preserve the properties of nonnegativity $\Pi_k({\bf p})\geq0$ and normalization $\sum_{k=1}^n\Pi_k({\bf p})=1$. The simple example of the  nonlinear map is given by the rational function of the form
\begin{equation}\label{eqq2}
\Pi^{(s)}_k({\bf p})=\frac{p_k^s}{\sum_{k=n_1}^{n_2}p^s_k},\quad 1\leq k=n_1, \, n_1+1,\ldots, n_2\leq n.
\end{equation}
Such map for $s=1$ gives, for example, conditional probability distribution. In fact,if a joint probability distribution $P(k,j)$ is written in the form of probability vector \\
${\bf p}=(P(1,1),P(1,2),\ldots,P(1,n),P(2,1),\ldots,P(n,m))$ the Bayes formula for conditional probability
\[{\bf p}\mapsto P(k|j)=\frac{P(k,j)}{\sum_{k=1}^n P(k,j)}\]
has the form (\ref{eqq2}) with choosing corresponding indices. Particular case of this map takes place for $n_1=1, \,n_2=n$. For example, if $s=2$ one has the map
\begin{equation}\label{eqq3}
{\bf p}\mapsto\vec\Pi^{(2)}({\bf p})=\Large(\sum_{k=1}^n p_k^2\Large)^{-1}(p_1^2,p_2^2,\ldots,p_n^2).
\end{equation}
Such maps can be considered as examples of nonlinear classical channels. In quantum case we define the nonlinear map of density $n\times n$ - matrix $\rho$ onto density $m\times m$ - matrix $R$ (i.e. $\rho_{k j}\mapsto R_{\alpha\beta}(\rho))$ preserving the properties of density matrices $R^\dag=R,\,\mbox{Tr}R=1,\,R\geq0$. The case of linear map of density matrices is the particular case of the map under discussion. For example, the positive linear map \cite{Stringp,Sud61} given in the form $R_{\alpha\beta}(\rho)=\sum_{k,j=1}^n B_{\alpha\beta,k j}\rho_{kj} $ and quantum channels corresponding to completely positive maps of the density matrix play important role in studying the quantum correlations in composite systems like entanglement phenomenon. The properties of the linear map positivity or complete positivity are coded by the properties of the matrix $B_{\alpha\beta,kj}$ \cite{Sud61}. The nonlinear maps of the density matrices which we call "nonlinear quantum channels" are characterized by the functions $R_{\alpha\beta}(\rho_{k j})$. One of the simple examples corresponding to example of probability vector transform (\ref{eqq2}) reads
\begin{equation}\label{eqq4}
R=\rho^s\frac{1}{\mbox{Tr}\rho^s},\quad s=2,3,\ldots,\infty.
\end{equation}
The map provides the new density matrix with larger purity which in generic case for $s\rightarrow\infty$ gives the pure state. The map can create entanglement, e.g. for two-qubit $X$-states.
An analog of classical Bayes formula for conditional probability distribution given by nonlinear map (\ref{eqq2}) for the matrix $\rho_{k j}$ $(k,j=1,2,\ldots,n)$ has the form
\begin{equation}\label{eqq5}
\rho_{k j}\mapsto R^{(m)}_{k'j'}=\frac{\rho_{k' j'}}{\sum_{k=1}^m\rho_{k k}}, k',j'=1,2,\ldots, m< n.
\end{equation}
The nonlinear positive map can be given in the form of map of the qudit tomogram. It is known (see, e.g.  \cite{MarmoManko}) that density matrix of arbitrary qudit system state with density matrix $\rho(1,2,\ldots,N)$ with $N$ subsystems is described by the tomographic probability distribution (qudit tomogram) which determines the density matrix. The probability vector $\vec w(u)$ corresponding to the density matrix has the vector components depending on unitary matrix $u$ and has the form
\begin{equation}\label{eqq6}
\vec w(u)=|u u_o|^2\vec\rho.
\end{equation}
Here $\vec\rho$ is the vector which has components equal to eigenvalues of the density matrix. The unitary matrix $u_0$ has as the columns the corresponding eigenvectors of the density matrix.
Using the expression (\ref{eqq6}) for tomogram of any qudit system state we formulated the general statement. Let us consider first case where the tomographic probability vector $\vec w(u)$ has $n=N\cdot M$ components. Then applying Eq.(\ref{eq.22}) to the vector we have inequality
\begin{equation}\label{eqq7}
-|u_0|^2\vec\rho\ln|u u_o|^2\vec\rho\leq-M_{12}|u u_0|^2\vec\rho\ln(M_{12}|u u_0|^2\vec\rho)-M_{21}|u u_0|^2\vec\rho\ln(M_{21}|u u_0|^2\vec\rho).
\end{equation}
This inequality is valid for tomogram of composite or noncomposite qudit system. The system is described by density matrix with eigenvalues providing $N\cdot M$ - vector $\vec\rho$ and corresponding eigenvectors combined into unitary matrix $u_0$. If the number $n\neq N\cdot M$ we introduce the vector with $n'= N\cdot M$ components where $n'=n+s$ adding $s$ extra zero components to the vector $\vec\rho$. The orthostohastic matrix $|u u_0|^2$ is extended also and replaced by the marix
\[\left(
    \begin{array}{cc}
      |u u_0|^2 & 0 \\
      0 & 1_s \\
    \end{array}
  \right)
     \]
where $1_s$ is $s\times s$-matrix.

Any map of density matrix, i.e. $\vec\rho\mapsto\vec\rho'$, $u_0\mapsto u_0'$ provides the map of the tomographic vector $\vec w(u)$. The rational map of density matrix (\ref{eqq4}) is equivalent to the map of probability vector $\vec\rho$ given by (\ref{eqq2}) with $n_1=1,\, n_2=n$. Thus, any linear or nonlinear map od the probability vector $\vec\rho$ which has the components equal to the density matrix eigenvalues yields the nonlinear positive map of the density matrix. Other nonlinear positive maps can be associated with change of the unitary matrix $u_0$, e.g. by means of linear map $u_0\mapsto u_0'=T u_0$, where $T$ is the unitary transform of the eigenvectors of the density matrix $\rho(1,2,\ldots,N)$. One has the entropic inequalities for the probability vector (\ref{eqq2}) $(n_1=1,\,n=n_2)$
\begin{equation}\label{eqq7}
-\sum_{k=1}^n\Pi_k^{(s+1)}({\bf p})\ln\Pi_k^{(s+1)}({\bf p})\leq-\sum_{k=1}^n\Pi_k^{(s)}({\bf p})\Pi_k^{(s)}({\bf p})\leq-{\bf p}\ln{\bf p}, \quad s=1,2,3,\ldots
\end{equation}
and for von Neumann entropy of the density matrix $R$ (\ref{eqq4}) one has
\begin{equation}\label{eqq8}
-\mbox{Tr}\Large[\large(\rho^s\frac{1}{\mbox{Tr}\rho^s}\large)\ln\large(\rho^s\frac{1}{\mbox{Tr}\rho^s}\large)\geq
-\mbox{Tr}\Large[\large(\rho^{s+1}\frac{1}{\mbox{Tr}\rho^{s+1}}\large)\ln\large(\rho^{s+1}\frac{1}{\mbox{Tr}\rho^{s+1}}
\large).
\end{equation}
One can introduce the positive map as the convex sum of the terms $\rho^s/\mbox{Tr}\rho^s$, i.e.
\[
  \rho\Rightarrow R=\sum_s p_s\large(\rho^s/\mbox{Tr}\rho^s\large), \quad 0\leq p_s\leq 1,\quad \sum_s p_s=1.
\]
The map gives the example of nonlinear channel. For composite bipartite systems the unitary transform which makes from the entangled states, separable states can be given by matrix $A$ such that
\begin{equation}\label{eqq9}
A=u_{01}\otimes u_{02},
\end{equation}
where the matrices $u_{ok}$ are unitary local transform matrices. For example, if the bipartite system with density matrix $\rho(1,2)$ has the eigenvectors providing unitary matrix $u_0$ such that
\begin{equation}\label{eqq10}
\rho(1,2)=u_0\rho_d u^+_0, \quad \rho_d=\left(
                                          \begin{array}{cccc}
                                            \rho_1 & 0 & 0 & 0 \\
                                            0& \rho_2 & 0 & o \\
                                            \ldots & \ldots & \ldots & \ldots \\
                                            0 & 0 & 0 & \rho_n \\
                                          \end{array}
                                        \right)
\end{equation}
any unitary matrix $A$ of the form
\begin{equation}\label{eqq11}
A=(u_{01}\otimes u_{02})u_0^+
\end{equation}
gives the tomographic vector
\begin{equation}\label{eqq12}
\vec w_A(u)=|u(u_{01}\otimes u_{02})|^2\vec\rho
\end{equation}
which is the tomogram of separable state. The applied transform depends on the state.

\section*{Conclusion}
To conclude we formulate main results of our work. We obtain new inequalities for both probability vectors and density matrices. These inequalities are analogs of known subadditivity conditions which are valid for composite systems but the inequalities are shown to be valid for arbitrary probability vectors and arbitrary density matrix including the case of systems without subsystems. We discussed the positive nonlinear maps of probability vectors and density matrices. The nonlinear maps can be used to create entangled states from the separable states. We considered explicitly the examples of the density matrix in six-dimensional Hilbert space which can be identified either with qubit-qutrit composite system state or with the state of single qudit with $j=5/2$. It is worthy to note that one can introduce Bell inequalities for noncomposite system. Also one can study the violation of the inequalities. The Bell inequality for the qudit with $j=3/2$ has the form of the inequality for two qubit system. The entangled states for the qudit ($j=3/2$) are the states for which equality for density matrix of the state in the form of separability condition is not valid. The Bell inequality can be violated. This problem will be considered in future publication.


\end{document}